\documentstyle[aps,prl,epsfig]{revtex}
\tightenlines

\newcommand{\be}{\begin{equation}}
\newcommand{\ee}{\end{equation}}

\begin{document}

\title{\Large{\bf Magnetic ordering of itinerant systems in modified mean field theory: antiferromagnetism}}

\vskip0.5cm 

\author{ J. Mizia,  G. G\'{o}rski, K. Kucab }

\address{Institute of Physics, University of Rzesz\'{o}w, ulica Rejtana 16A, \\
35-958 Rzesz\'{o}w, Poland\\}

\maketitle

\vskip0.5cm 


\begin{abstract}

\noindent
 We analyze the itinerant model for antiferromagnetism, which was developed previously by Plischke, Mattis, Brouers and Mizia. In this model we include both; single-site and 
two-site electron correlations. Including additionally band degeneration into the model allows for considering intra-atomic exchange interactions in the Hamiltonian. The modified 
Hartree-Fock approximation for the two-site interactions is used. This approximation gives the spin band narrowing, which is the same for both spin directions and affects 
possibility of antiferromagnetic ordering. We use the Green function technique and CPA decoupling. This allows for the change in shape of the spin bands, described by the 
correlation factors, which decreases kinetic energy of the system. The effective Hartree field and the gain in kinetic energy due to the on-site and inter-site correlation factors drive 
the antiferromagnetism.

\end{abstract}

\vskip0.5cm



\vskip1.5 cm
\noindent {\Large {\bf 1. Introduction} }

\vskip0.5cm 

The basic model for magnetic ordering of itinerant electrons in solids is the Hubbard model \cite{1}. In the mean-field approximation the Hubbard model leads to the well-known 
Stoner model for magnetism. Extension of the Hubbard model is the model, which in addition to the Coulomb repulsion $U_0$ includes also Coulomb-"exchange" interaction 
$J=(i\lambda,j\nu\mid 1/r\mid j\nu,i\lambda)$. The solutions of the 3-dimensional Hubbard model did not yield the exchange field small enough to be justified physically, therefore 
to obtain the magnetic state Hirsch (\cite{2,3}) and others (see e.g. \cite{4}) included also the nearest-neighbor inter-site interactions; $(i,j\mid 1/r\mid k,l)$, where $(k,l)=(i,j)$ and 
$i$, $j$   are the nearest-neighbor lattice sites. In our model we introduce into Hamiltonian the on-site Coulomb repulsion $U_0$ single-site two sub-band interactions and also the 
inter-site nearest-neighbor interactions. Analyzing the inter-site correlations we deal with probabilities defined for the product of four operators, in which two-operator product is 
replaced by its stochastic value. We have interpreted the stochastic value of two-operator product as the product of probabilities for electron transfer between two neighboring 
atoms, which is the product of probability that site i is empty for electron witch spin $+ \sigma$; $1 - n^\sigma$ , and the probability of site $j$ being occupied; $n^\sigma$. This 
interpretation makes the connection of standard probabilities used in the CPA method with Hirsch's average bond energy for spin $\sigma$; $I_\sigma=n^\sigma(1-n^\sigma)$. 
The connection allows us to expand our model to more realistic cases of the weak inter-site interaction being accompanied by the weak or strong single-site Coulomb interaction.

\vskip1.0cm 

\noindent {\Large {\bf 2. Hamiltonian and the Coherent Potential Approximation
}} 

\vskip0.5cm 

The Hamiltonian for one degenerate band can be written in the form, which we already previously introduced (see \cite{5}) 
\be 
H =  - \sum\limits_{\scriptstyle ij,\lambda \nu  \hfill \atop 
  \scriptstyle \,\,\,\,\sigma  \hfill} {t_{ij} \left( {c_{i\lambda \sigma }^ +  c_{j\nu \sigma }  + h.c} \right)}  - \mu _0 \sum\limits_{i,\lambda ,\sigma } {\hat n_{i\lambda \sigma } }  + 
\sum\limits_{\scriptstyle \,\,\,\,ijkl, \hfill \atop 
  {\scriptstyle \lambda \nu \varpi \vartheta  \hfill \atop 
  \scriptstyle \,\, \sigma ,\sigma ' \hfill}} {\left( {i\lambda ,j\nu |1/r|k\vartheta ,l\varpi } \right)c_{i\lambda \sigma }^ +  c_{j\nu \sigma '}^ +  c_{l\varpi \sigma '} c_{k\vartheta \sigma 
} } ,
\label{1}
\ee 

\noindent
where $t_{ij}$ is the nearest neighbors hopping integral, $\mu _0$ is the chemical potential, $c_{i\lambda \sigma }^ +  \left( {c_{i\lambda \sigma } } \right)$ creates (destroys) an 
electron of spin $\sigma$   in a Wannier orbital $\lambda$   on the $i$-th lattice site, the indices $\lambda ,\nu ,\varpi ,\vartheta$ numerate the sub-bands in the degenerated single 
band. 
Taking into account in Hamiltonian (\ref{1}) only single-site and two-site interactions, and also single sub-band and two sub-band interactions we obtain and retain the following 
matrix elements; \\
\begin{itemize}
 \item single-site, single sub-band interaction\\
 $
U_0  = \left( {i\lambda ,i\lambda |1/r|i\lambda ,i\lambda } \right),
$
 \item single-site (subscript "in"), two sub-band interactions (for $\lambda  \ne \nu $ )\\
 $V_{in}  = \left( {i\lambda ,i\nu |1/r|i\lambda ,i\nu } \right)$,\,\,\,\, $J_{in}  = \left( {i\lambda ,i\nu |1/r|i\nu ,i\lambda } \right)$, $J'_{in}  = \left( {i\lambda ,i\lambda |1/r|i\nu ,i\nu } 
\right),$
 \item two-site interactions\\
 $V_0  = \left( {i\lambda ,j\nu |1/r|i\lambda ,j\nu } \right)$,\,\,\,\,\, $J_0  = \left( {i\lambda ,j\nu |1/r|j\nu ,i\lambda } \right)$,\\ $J'_0  = \left( {i\lambda ,i\lambda |1/r|j\nu ,j\nu } 
\right)$,\,\,\,\,\, $\Delta t_0  = \left( {i\lambda ,i\lambda |1/r|j\nu ,i\lambda } \right).$
 \end{itemize}
 Keeping these interactions in Hamiltonian (\ref{1}) we obtain \\
 $$
H =  - t_0 \sum\limits_{ < ij > \sigma } {\left( {c_{i\sigma }^ +  c_{j\sigma }  + h.c.} \right)}  - \mu _0 \sum\limits_i {n_i }  + U\sum\limits_{i\sigma } {n_{i\sigma } n_{i - \sigma } } 
 + V\sum\limits_{ < ij > } {n_i n_j } $$
 \be
  + \sum\limits_{ < ij > \sigma } {\Delta t\left( {c_{i\sigma }^ +  c_{j\sigma }  + h.c.} \right)\left( {n_{i - \sigma }  + n_{j - \sigma } } \right)}+ J\sum\limits_{ < ij > \sigma \sigma '} 
{c_{i\sigma }^ +  c_{j\sigma '}^ +  c_{i\sigma '} c_{j\sigma }  + } J'\sum\limits_{ < ij > \sigma } {c_{i\sigma }^ +  c_{i - \sigma }^ +  c_{j - \sigma } c_{j\sigma } }, 
\label{2}
\ee
\noindent where in the degenerated band model in the Hartree-Fock approximation we have 
\be
U = U_0  + (p - 1)(J_{in}  + J'_{in}  + V_{in} ),
\label{3}
\ee
\noindent and $p$  is the number of degenerated sub-bands in the band.
The effective inter-atomic constants $J$, $J'$, $V$, $\Delta t$   for the degenerate band are given by  $pJ_0$, $pJ'_0$, $pV_0$, $p\Delta t_0$  respectively. 
We will apply mean field approximation to the inter-atomic weak interactions $J$, $J'$, $V$ and  $\Delta t$. 
After these approximations the Hamiltonian (\ref{2}) will take on the following form
\be
  H =  - \sum\limits_{\scriptstyle ij \hfill \atop 
  \scriptstyle \sigma  \hfill} {t_{ij}^\sigma  \left( {c_{i\sigma }^ +  c_{j\sigma }  + h.c.} \right)}  - \sum\limits_i {\mu _0 } n_i   + \sum\limits_{i\sigma } {M^\sigma  n_{i\sigma } }  + 
U\sum\limits_{i\sigma } {n_{i\sigma } n_{i - \sigma } },  
\label{4}
\ee
\noindent where the mean (molecular) field is given by
\be
M^\sigma   = J\bar n_j^{ - \sigma }  + Vn + \Delta t\left\langle {c_{i - \sigma }^ +  c_{j - \sigma } } \right\rangle 
\label{5}
\ee
\noindent and the spin dependent hopping integral $t_{ij}^\sigma$  is expressed as
\be
t_{ij}^\sigma   = t_0  - \Delta t\left( {\bar n_{i - \sigma }  + \bar n_{j - \sigma } } \right) + \left( {J + J'} \right)I_{ij}^{ - \sigma }  + \left( {J - V} \right)I_{ij}^\sigma , 
\label{6}
\ee
\noindent where $\bar n_{i\sigma } $   is the average number of electrons with spin  $\sigma$ on the$i$-th lattice site, and the parameter $I_{ij}^\sigma  {\rm  }$  is defined as 
$I_{ij}^\sigma   = \left\langle {c_{i\sigma }^ +  c_{j\sigma } } \right\rangle$. 

In the case of ferromagnetism parameter $I_{ij}^\sigma  {\rm  }$  was spin dependent (see e.g. \cite{2,5}). For the antiferromagnetism indices $i,j$ belong to the neighboring 
sublattices $\alpha, \beta$  with opposite magnetic moments. In the result $I_{ij}^\sigma   = I_{ij}^{ - \sigma }  = I_{AF} $  is spin independent.

\vskip1.0cm 

\noindent {\Large {\bf 3.  Application to antiferromagnetism }} 

\vskip0.5cm 

For the antiferromagnetism we will use the diagonalization of Plischke and Mattis \cite{6}, Brouers \cite{7} and Mizia \cite{8}. The crystal lattice will be divided into two 
interpenetrating sublattices $\alpha, \beta$, with the average electron numbers equal to 
\be
n_\alpha ^{ \pm \sigma }  = \frac{{n \pm m}}{2},\;\quad n_\beta ^{ \pm \sigma }  = \frac{{n \mp m}}{2},
\label{7}
\ee
\noindent where magnetization $m = n_\alpha ^\sigma   - n_\alpha ^{ - \sigma }  = n_\beta ^{ - \sigma }  - n_\beta ^\sigma$. We use mean field approximation for the interactions 
between the neighboring lattice sites. For the stronger on-site interaction we use the coherent potential approximation (CPA). To describe the antiferromagnetic state we write the 
model Hamiltonian (\ref{2}) taking into account the two interpenetrating magnetic sublattices  , $\alpha, \beta$, with magnetic moments given by Eq. (\ref{7}) 
\be
H =  - \sum\limits_{i \in \alpha } {\sum\limits_{\scriptstyle j \in \beta  \hfill \atop 
  \scriptstyle \sigma  \hfill} {t_{ij}^\sigma  \left( {c_{i\sigma }^ +  c_{j\sigma }  + h.c.} \right)} }  - \sum\limits_{i\sigma } {\left( {\mu _0  - M^\sigma  } \right)} \,\;n_{i\sigma }  + 
\sum\limits_{\scriptstyle i \in \alpha  \hfill \atop 
  \scriptstyle \sigma  \hfill} {Un_{i\sigma } \;n_{i - \sigma } }  + \sum\limits_{\scriptstyle i \in \beta  \hfill \atop 
  \scriptstyle \sigma  \hfill} {Un_{i\sigma } \;n_{i - \sigma } }. 
\label{8}
\ee

Next we use the main idea of the CPA formalism \cite{9} and split the above stochastic Hamiltonian into the homogeneous part 
\be
H_0  =  - \sum\limits_{i \in \alpha } {\sum\limits_{\scriptstyle j \in \beta  \hfill \atop 
  \scriptstyle \sigma  \hfill} {t_{ij}^\sigma  \left( {c_{i\sigma }^ +  c_{j\sigma }  + h.c.} \right)} }  - \sum\limits_{i\sigma } {\mu \;} n_{i\sigma }  + \sum\limits_{\scriptstyle i \in 
\alpha  \hfill \atop 
  \scriptstyle \sigma  \hfill} {\Sigma _\alpha ^\sigma  \;n_{i\sigma } }  + \sum\limits_{\scriptstyle i \in \beta  \hfill \atop 
  \scriptstyle \sigma  \hfill} {\;\Sigma _\beta ^\sigma  n_{i\sigma } } 
\label{9}
\ee
and the stochastic part 
\be
H_I  = \sum\limits_{\scriptstyle i \in \alpha  \hfill \atop 
  \scriptstyle \sigma  \hfill} {(\tilde V_{i\alpha }^\sigma   - \Sigma _\alpha ^\sigma  )\;n_{i\sigma } }  + \sum\limits_{\scriptstyle i \in \beta  \hfill \atop 
  \scriptstyle \sigma  \hfill} {\;(\tilde V_{i\beta }^\sigma   - \Sigma _\beta ^\sigma  )n_{i\sigma } }, 
\label{10}
\ee
where $\tilde V_i^\sigma$  is the stochastic potential given by 
\be
\tilde V_{ix}^\sigma   = \left\{ \begin{array}{l}
 Jn_x^\sigma  {\rm           }P_{1x}^\sigma   = 1 - n_x^{ - \sigma }  \\ 
 U + Jn_x^\sigma  {\rm       }P_{2x}^\sigma   = n_x^{ - \sigma }  \\ 
 \end{array} \right.
\label{11}
\ee
$\mu  = \mu _0  - Vn - \Delta tI_{AF}$  and  $\Sigma _x^\sigma$ are the self-energies on sites $x = \alpha (\beta )$  for electrons with spin $\sigma$.
Using Hamiltonian (\ref{9}) in the equations of motion for Green functions we obtain the following equations in the momentum space
$$
\left[ {\begin{array}{*{20}c}
   {\varepsilon  + \mu  - \Sigma _\alpha ^\sigma  } & { - \varepsilon _k }  \\
   { - \varepsilon _k } & {\varepsilon  + \mu  - \Sigma _\beta ^\sigma  }  \\
\end{array}} \right]\left[ {\begin{array}{*{20}c}
   {G_\sigma ^{\alpha \alpha } (k,\varepsilon )} & {G_\sigma ^{\alpha \beta } (k,\varepsilon )}  \\
   {G_\sigma ^{\beta \alpha } (k,\varepsilon )} & {G_\sigma ^{\beta \beta } (k,\varepsilon )}  \\
\end{array}} \right] = {\bf \hat 1},
$$
where ${\bf \hat 1}$  is the unity matrix.

Next we use these Green functions in the relations for electron numbers 
\be
n_\alpha ^{ \pm \sigma }  = \frac{1}{N}\sum\limits_k {\int\limits_{ - \infty }^\infty  {f\left( \varepsilon  \right)\left( { - \frac{1}{\pi }} \right)} {\mathop{\rm Im}\nolimits} \left[ 
{G_{ \pm \sigma }^{\alpha \alpha } \left( {\varepsilon ,k} \right)} \right]} d\varepsilon, 
\label{12}
\ee
where for $x = \alpha \left( \beta  \right)$  we have respectively
$$
G_\sigma ^{xx} \left( {\varepsilon ,k} \right) = \frac{1}{2}\sqrt {\frac{{\varepsilon  + \mu  - \Sigma _{\beta \left( \alpha  \right)}^\sigma  }}{{\varepsilon  + \mu  - \Sigma _{\alpha 
\left( \beta  \right)}^\sigma  }}} \left[ {G^0 \left( {k,\varepsilon _{eff} } \right) - G^0 \left( {k, - \varepsilon _{eff} } \right)} \right],
$$

$$
\varepsilon _{eff}  = \sqrt {\left( {\varepsilon  + \mu  - \Sigma _\alpha ^\sigma  } \right)\left( {\varepsilon  + \mu  - \Sigma _\beta ^\sigma  } \right)} ,\quad G^0 \left( {k,\varepsilon } 
\right) = \frac{1}{{\varepsilon  - \varepsilon _k }}.
$$
If we use these results and changing summation over "k" to integration, and inserting for self-energies their first order approximation
\be
\Sigma _{\alpha (\beta )}  = F^{AF} \frac{n}{2} \mp F^{AF} \frac{m}{2} = F^{AF} \frac{n}{2} \mp \Delta, 
\label{13}
\ee
we obtain
$$
n_\alpha ^{ \pm \sigma }  = \frac{1}{4}\int\limits_{ - \infty }^\infty  {\rho (\varepsilon )\frac{\varepsilon }{E}} \left[ {S^{ \pm \sigma } f\left( {E - \mu } \right) + S^{ \mp \sigma } 
f\left( { - E - \mu } \right)} \right]d\varepsilon, 
$$
where $E = \sqrt {\varepsilon ^2  + \Delta ^2 }$, $S^{ \pm \sigma }  = \sqrt {\frac{{E \pm \Delta }}{{E \mp \Delta }}}$ and $f(E)$ is the Fermi distribution function.
The antiferromagnetic energy gap $\Delta$  is given by the expression; $\Delta  = F^{AF} \frac{m}{2}$, in which  $F^{AF}$ is the value of the effective exchange interaction 
$F^{AF}  = U - zJ$. 
It can be seen that contrary to the case of ferromagnetism (see \cite{5}) the inter-atomic exchange interaction, $J$, in effective field $F^{AF}$  is opposing the antiferromagnetism. 
The antiferromagnetic moment per atom (in Bohr's magnetons) will be given now by the following expression
\be
m = \frac{1}{{2N}}\sum\limits_k {\left( {P_k^{ + \sigma }  - P_k^{ - \sigma } } \right)\left[ {f(E_k ) - f( - E_k )} \right]}, 
\label{14}
\ee
where $P_k^{ \pm \sigma }$  is the occupation probability of state $(k, \pm \sigma )$, $E_k  = \sqrt {\varepsilon _k^2  + \Delta ^2 }$,   and
\be
P_k^{ \pm \sigma }  = \frac{{\varepsilon _k^{ \pm \sigma } }}{{E_k }}.
\label{15}
\ee
The effective electron dispersion relation $\varepsilon _k^{ \pm \sigma }$  used above has the following form 
\be
\varepsilon _k^{ \pm \sigma }  = \varepsilon _k S_k^{ \pm \sigma }, 
\label{16}
\ee
where $$\varepsilon _k  = \varepsilon _k^0 L_{AF} \left( {n,m} \right),\quad S_k^{ \pm \sigma }  = \sqrt {\frac{{E_k  \mp \;\Delta }}{{E_k  \pm \;\Delta }}},$$
$$L_{AF} \left( {n,m} \right) = 1 - n\frac{{\Delta t}}{{t_0 }} + \frac{1}{{t_0 }}\left( { - 2J - J' + V} \right)I_{AF} \left( m \right).$$
The chemical potential $\mu$  may be determined from a carrier concentration $n$ on the base of the equation 
\be
n = \frac{1}{{2N}}\sum\limits_k {\left( {P_k^{ + \sigma }  + P_k^{ - \sigma } } \right)\left[ {f(E_k ) + f( - E_k )} \right]}. 
\label{17}
\ee
Inter-atomic interactions $J$, $J'$, $V$   and $\Delta t$  used in this model change the effective half bandwidth according to the relation, where $D = D_0 L_{AF} (n,m)$  is a factor 
of the bandwidth change.
For the antiferromagnetic ordering the change of width of the band is independent from spin direction, since
\be
I_{AF} \left( m \right) = \frac{1}{N}\sum\limits_k {\frac{{\varepsilon _k^2 }}{{DE_k }}\left[ {f(E_k ) - f( - E_k )} \right]}. 
\label{18}
\ee
Inserting Eq. (\ref{16}) and Eq. (\ref{15}) to Eq. (\ref{14}) we obtain
\be
1 =  - F^{AF} \frac{1}{N}\sum\limits_k {\frac{1}{{2E_k }}\left[ {f(E_k ) - f( - E_k )} \right]}. 
\label{19}
\ee

\vskip1.0cm 

\noindent {\Large {\bf 4. Numerical examples}} 

\vskip0.5cm 

We calculated the critical values of interaction $F_{cr}^{AF}$  from Eq. (\ref{19}) for different values of assisted hopping  $\Delta t$ assuming that the other inter-site 
interactions are equal zero. We used constant density of states. The assisted hopping  $\alpha  = \Delta t/t$ corresponds to the solid line in Fig. 1. Different curves calculated from 
complete Eq. (\ref{19}) for different values of assisted hopping $\alpha$  are also shown in Fig. 1. One can see that with growing $\alpha$  the critical exchange constant 
necessary for AF is decreasing for small hole concentration. The decrease is even stronger in the split band limit at $n \to 1$ . This effect could possibly explain the existence of 
the AF at the almost full subband in the high temperature superconductors. There is one limitation in here; that these are \em s-wave \rm $\left( {s_0  + s_{x^2  + y^2 } } \right)$ 
superconductors, since the assisted hopping contributes only to the superconductivity of this type of symmetry (see \cite{10}  for more detailed analysis of this problem).

\begin{figure}[t]
\begin{center}
\epsfig{file=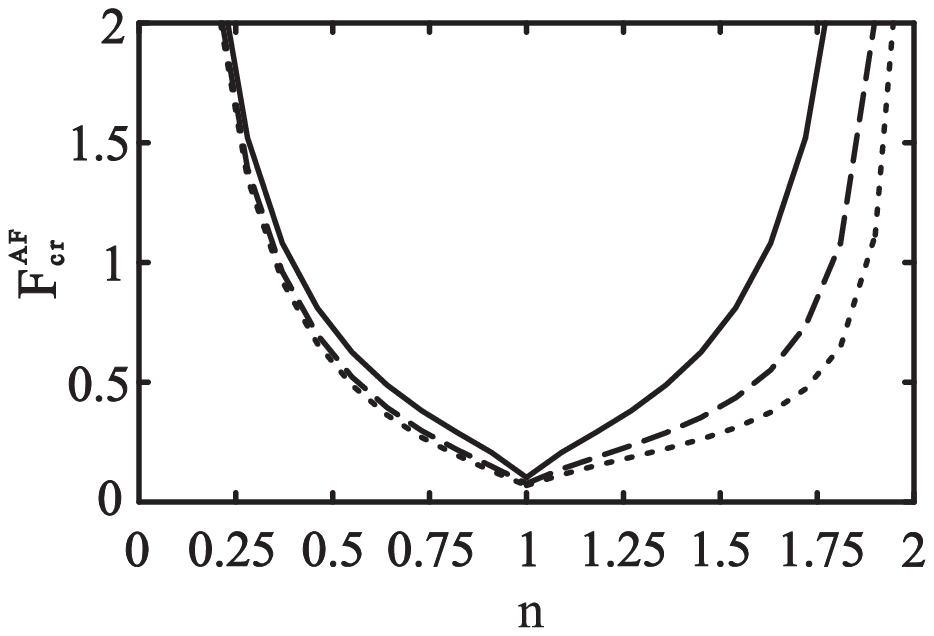,width=0.45\hsize}
 \par\vspace{1.5ex}\makebox[0.5\hsize]
    {\small FIG. 1}
	\end{center}
	\vskip0.5 cm
	\caption{Dependence of critical field on carrier concentration $F_{cr}^{AF}$ for constant density of states with assisted hopping only; $\alpha=0$ - solid line, $\alpha=0.3$ - 
dashed line, $\alpha=0.4$ - dotted line.}
\vskip0.5 cm
\end{figure}


\begin{references}   

\bibitem{1} J. Hubbard, Proc. Roy. Soc. A{\bf 276}, 238 (1963). 
\bibitem{2} J.C. Amadon, J.E. Hirsch, Phys. Rev. B{\bf 54}, 6364 (1996). 
\bibitem{3} J.E. Hirsch, Phys. Rev. B{\bf 59}, 6256 (1999). 
\bibitem{4} R. Strack and D. Vollhardt, Phys. Rev. Lett. {\bf 72}, 3425 (1994). 
\bibitem{5} J. Mizia, G. G\'{o}rski and K. Kucab, Physica B{\bf 325}, 106 (2003).
\bibitem{6} M. Plischke and D. Mattis, Phys. Rev. B{\bf 7}, 2430 (1973).
\bibitem{7} F. Brouers, J. Giner and J. Van der Rest, J. Phys. F{\bf 4}, 214 (1974).
\bibitem{8} J. Mizia, phys. stat. sol. (b){\bf  84}, 449 (1977).
\bibitem{9} B. Velick\'{y}, S. Kirkpatrick and H. Ehrenreich, Phys. Rev. {\bf 175}, 747 (1968).
\bibitem{10} G. G\'{o}rski and J. Mizia, phys. stat. sol. (b){\bf 215}, 1083 (1999).

\end{references}
\end{document}